\documentclass[floats]{iopart}
\usepackage{epsfig}

\def\be{\begin{equation}}
\def\bea{\begin{eqnarray}}
\def\eea{\end{eqnarray}}
\def\ee{\end{equation}}
\def\bi{\begin{itemize}}
\def\ei{\end{itemize}}

\def\bn{\begin{enumerate}}
\def\en{\end{enumerate}}

\def\be{\begin{equation}}
\def\ee{\end{equation}}
\def\bea{\begin{eqnarray}}
\def\eea{\end{eqnarray}}

\def\beq{\begin{equation}}
\def\eeq{\end{equation}}

\begin{document}
\title{Detection of periodic gravitational wave sources by Hough transform in the  $f$ vs $ \dot{f}$ plane}
\author{ {\bf F.Antonucci, P.Astone, S.D'Antonio, S.Frasca, C.Palomba}\\
INFN Rome ``La Sapienza'', INFN Rome 2 ``Tor Vergata'' and 
University ``La Sapienza'', Rome}
\date{}
\begin{abstract}
In the hierarchical search for periodic sources of gravitational waves, 
the candidate selection, in the
incoherent step, can be performed with Hough transform procedures.
In this paper we analyze the problem of sensitivity loss due to discretization 
of the parameters space vs computing
cost, comparing the properties of the sky Hough procedure with those
of a new frequency Hough, which is
based on a transformation from the {\it time - observed frequency} plane to 
the {\it source frequency - spin down} plane. Results on simulated peakmaps
suggest various advantages in favor of the use of the frequency Hough.
The ones which show up to really make the difference are 1) the possibility to
enhance the frequency resolution without relevantly affecting the 
computing cost. This reduces the digitization effects; 2) the excess of candidates 
due to local disturbances in some places of the sky map. They
 do not affect the new analysis because each map is constructed for
only one position in the sky. \\
Pacs. numbers: 04.80Nn,07.05Kf,97.60Jd
\end{abstract}
\section{Introduction}
One of the main goals of the search for periodic isolated sources of 
gravitational waves (g.w.) is  
to perform all sky surveys, based on
``blind searches'', where the source parameters are
unknown. In this case
hierarchical procedures are applied, 
based on a sequence of increasing resolution steps.
In this paper we study in details the problem of sensitivity loss 
due to discretization of parameters and to the needs to limit 
the computing cost, with Hough
procedures. In particular, we propose and study the characteristics of
a frequency Hough procedure, designed mainly to reduce the discretization 
problem,
and we  compare it with the sky Hough procedure,
which is actually used in the Virgo collaboration. \\
The paper is organized as follows: in Sect. 2 we present the basic scheme
of the Rome hierarchical procedure, based on the main 
idea of coincidences among subsets of data;
in Sect. 3 we discuss the limits due to digitization of the sky Hough 
procedure; 
in Sects. 4, 5 we present the new frequency Hough procedure, 
discussing details its implementation and its basic characteristics;
in Sect. 6 we present the study of amplitude losses due to digitization, 
and thus efficiencies,
for both the procedures. Conclusions and comments are given in Sect. 7.

\section{Scheme of the hierarchical procedure and Hough transforms} 
Hierarchical procedures, based on Hough transform algorithms, are
applied by various groups in the g.w. community. See, for example, references
 \cite{hough1,hough2}. There are various ways of implementing the 
hierarchical procedure and the Hough transform.
\subsection{Hough transform basic characteristics}
The Hough transform is a linear transform that is used to recognize the 
parameters of the analytical description of a curve from the position of 
some points on it. 
It operates on an ``image'' of points, in our case the peakmap in 
the time-frequency plane. For each peak of this map we increase a 
set of bins of a multi-dimensional histogram 
(in our case a two-dimensional histogram) 
defined on the parameters space, called the Hough map.
In the old procedure, the parameter space 
was the position of the source, i.e. the celestial sphere, and we fixed the 
spin down value for each Hough map. 
In the new one, the parameter space is the plane $f \,- \,\dot f$, and for 
each Hough map, we fix the position of the source.
The mapping (i.e. which points of the Hough map must be increased for a 
certain point in the peakmap) can be done in different ways: 
we use always what we call the ``biunivocal mapping'', i.e. a mapping in 
which every point in the Hough map derive from a single point of the 
peakmap at a given time. 
It is easy to demonstrate that in this case the mapping is also uniform, 
i.e. in the case of uniformly distributed random dots in the peakmap, 
the expected value of the Hough map $\mu$ is a constant 
(for all parameter value). 
This value, depending on the number N of the spectra of the peakmap and on 
the mapping, defines the ``noise'' of the map. 
It is binomially distributed with parameters N and $p=\mu/N$.

\subsection{Scheme of the hierarchical procedure}
\begin{figure}
\begin{center}
\includegraphics[width=6cm]{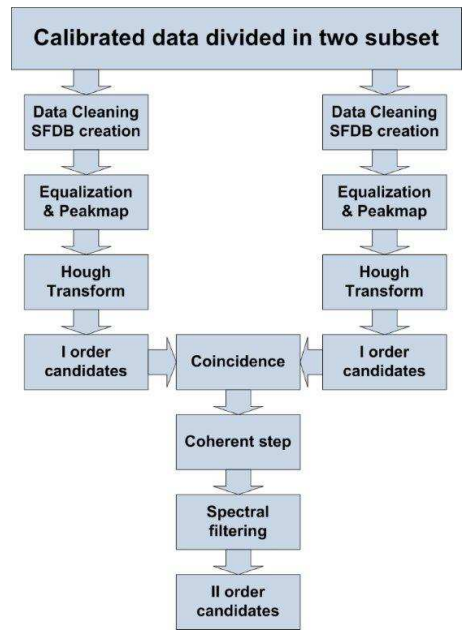}
\end{center}
\caption{\small \sf The basic scheme of the hierarchical procedure, used 
in the Virgo collaboration}  
\label{fig:schema}
\end{figure}

We will refer here to the Rome scheme, presently used in Virgo data.
Fig. \ref{fig:schema} shows the basic scheme of the Rome hierarchical procedure.
Details on the main aspects of the procedure 
are given in references \cite{piaqui,sergioqui,cristianoqui}.
After data cleaning (short time domain disturbances removal) and 
``Short FFTs Data Base'' (SFDB) creation,
peakmaps are computed, using a very refined auto-regressive algorithm 
to equalize the spectral data by an appropriate follow-up of 
the noise. Peakmaps are
frequency vs time maps, 
obtained from equalized spectra by selecting all the local maxima 
above a chosen threshold. An accurate 
cleaning of peakmaps, by removing known noise lines and the 
more persistent lines,
is needed and its implementation is critical for the next step analysis.
On the cleaned peakmaps, methods of peaks detection are applied. That is,
transformation from the input plane to the Hough plane, 
thresholding and first order candidates selection. Candidate parameters are 
defined by
source frequency, celestial coordinates, first spin-down parameter. 
The need for coincidences among candidates obtained in different 
subsets of data
(two in the scheme of Fig. \ref{fig:schema}) has been discussed in 
references \cite{c6c7,mg11_1,mg11_2}.
This method is very efficient to reduce the 
number of spurious candidates at a fixed threshold. 
Thus, for a given false alarm probability, we can lower the 
threshold -with respect to the choice of not doing
coincidences- gaining in detection efficiency. The method has a better 
efficiency when the data sets have similar sensitivities.
After the coincidence, the survived candidates are analyzed 
coherently with longer FFTs on corrected data. Then the spectral
filtering is used  to take into account the spread of the power in five 
bands, as explained in reference \cite{spectfilt}.
Finally, second order candidates are produced.
\section{The problem}
As stated before, the sky Hough method shows
amplitude losses, and thus loss of sensitivity, which are due to digitization
of parameters. 
This effect shows up mainly for
the complexity of the transform together with the need of reducing 
the computing cost:
\begin{itemize}
\item
the method is based on a transform  between  
the time-frequency peakmap 
and the celestial sphere. 
It is not simple for the non linearity of the mapping; 
\item
to reduce the computational effort, 
we need to use ``look-up tables'' which introduce further digitization errors; 
\item
to reduce the computational effort, fast algorithms have been developed,
which require the use of a rectangular grid to map the sky. Compared to the
``optimal'' (see later) grid, the rectangular one has
over-resolution in some regions of the sky. 
This leads also to a higher number of candidates.
\item
the use of the celestial map as the space to spot the candidates is very 
prone to artifacts, see \cite{c6c7}: some regions are always ``privileged'', that is 
they have a higher candidates number with respect to the expectation. The problem
arises because each Hough map is constructed over the whole sky.
\end{itemize}
Hence, it seemed important the study of alternative procedures.
Given the observation that most of the problems are related to the 
complexity of the transformation, we exploit the possibility of the use 
of a different but
simpler transformation. A part the simplicity of the transformation we obviously
need to study a procedure which is  less, or equivalently, computationally expensive. 
Therefore we studied a procedure which has a better, or equivalent,
sensitivity, at the same computational cost of the sky Hough.
\section{The frequency Hough}
The transformation we propose transforms the {\bf Time - observed frequency} 
plane
into the   {\bf Source frequency - Spin down} plane.
Let' s go into details.
If $f$ is the frequency
(Doppler corrected for a given sky direction), 
$f_0$ the source intrinsic frequency, $d=\dot f$ the first 
spin-down parameter, $t$ the time at the detector and $t_0$ a reference time,
we have that
\begin{equation}
f=f_0+ d\,(t-t_0)
\end{equation}
 a straight line in the Hough plane. We then get the following:
\begin{equation}
d= -\frac{f_0}{t-t_0} + \frac{f}{t-t_0}
\label{eq:uno}
\end{equation}
Each point in the input plane $(t-t_0,f)$, 
that is a peak in the Doppler shifted peakmap, is transformed into  a straight line 
in the Hough
$(f_0,d)$ plane, with slope $-1/(t-t_0)$. The slope depends on 
the choice of the reference time. If we choose $t_0$ equal to the
beginning time of the data we analyze, then the slope is always negative and inversely
proportional to the time gap. \\ This is the choice we have done here.
In addition, considering the width $\Delta f$ of the frequency bins in the 
input plane we notice that each peak is transformed into a stripe among 
two parallel straight lines

\begin{equation}
 -\frac{f_0}{t-t_0} +\frac{f - \Delta f/2}{t-t_0}
< d < -\frac{f_0}{t-t_0} +\frac{f+ \Delta f/2}{t-t_0}\label{eq:due}
\end{equation}

It is a linear transformation.
Now the input plane is obtained from the original peakmap
by correcting it for the Doppler shift due to the Earth revolution and rotation,
for each point in the sky grid we need to analyze. Thus ``time'' is the
time at the detector and ``frequency'' the observed frequency, after the Doppler
correction.  
But, as each SFDB is short enough to not be affected by a time-varying 
Doppler shift, then the Doppler effect 
removal from 
the original peakmap, obtained from the collection of all the 
SFDB data,
 reduces to a very simple ``shifting'' procedure of the 
peakmap bins. In the analysis scheme, this bins shift is part of the 
Hough procedure.\\
In the following, we give details on the construction of the map.
\subsection{The direct differential method}
The frequency Hough map is constructed using 
the ``direct differential method'',
as is done with the sky Hough.
With this method, instead of building directly the Hough map, one builds a
map that, if ``integrated'' (i.e. summed over bins from left to right),
gives the Hough map. 
This is important to minimize the number of floating point operations.
As already explained, for each sky position, the input peakmap is got 
from the original one
by shifting bins to correct for the Doppler effect.
The sky is sampled with a non uniform covering grid, which will be later
discussed.
Here we explain in detail the technique, by giving the sequence 
of operations: 
\begin{itemize}
\item
for each point in the sky grid,
for each coordinate in the input plane $(t-t_0,f)$ and
for each spin-down value $d$,
\end{itemize}
the map is incremented by 1 in the point
$$f_0=f - \frac{\Delta f}{2}-d \cdot (t-t_0)$$
and decremented by 1 in the point
$$f_0=f + \frac{\Delta f}{2}-d \cdot (t-t_0)$$.

Hence, for each sky position, a differential map is 
constructed. The sum of the bins along the frequency direction is then
performed to construct the final integral map. 
This two dimensional histogram is the frequency Hough map.
In the algorithm implementation we plan to divide the input peakmap into 
10 Hz bands, thus constructing, for each position in the sky, a different Hough map every 10 Hz. \\
In case there is the need to exploit higher order one spin down parameters,
one (or more) loop(s) has (have) to be added to the 
sequence of operations, to scan the discrete set
of values of the new parameter(s). This clearly influences the 
computing cost, but does not change the basics of the method. 
\section{Main characteristics: frequency resolution and sky grid}
Let's first discuss two peculiar
aspects of this new method, which are the basis of its appeal.
\subsection{Increasing the frequency resolution}
From the given analysis scheme, it is easy to 
see that the frequency resolution for the
estimation of the source frequency $f_0$ can be enhanced, with respect to the
binning frequency $\Delta f$, without relevantly affecting the computational
effort. In fact, the use of a resolution
\begin{equation} 
\Delta f_0=\Delta f/r 
\end{equation}
with $r >= 1$, affects only the size of the  
Hough map. This has a computational cost only when summing over the
bins to construct the integral map from the differential one. 
But we notice that the total cost of the construction 
of the Hough map is due to the construction of the differential map, dominated 
by the number of peaks in the peakmap and to the construction of the
integral map, dominated by the number of bins. The former, in all 
practical cases, is the one which dominates. \\
The possibility to enhance the frequency resolution
results to be, as will be shown in the next sections,
a very important peculiarity of the new method. It 
which enhances considerably the efficiency, by reducing the 
digitalization effect.
 The same in the sky Hough
procedure would have a relevant computational cost. 
Regarding the increasing of the spin down resolution, it would cost for
both the procedures:
the better the resolution in the spin down estimation
the higher is the number of loops of the procedures.
\subsection{The grid on the sky}
\begin{figure}
\begin{center}
\includegraphics[width=12cm]{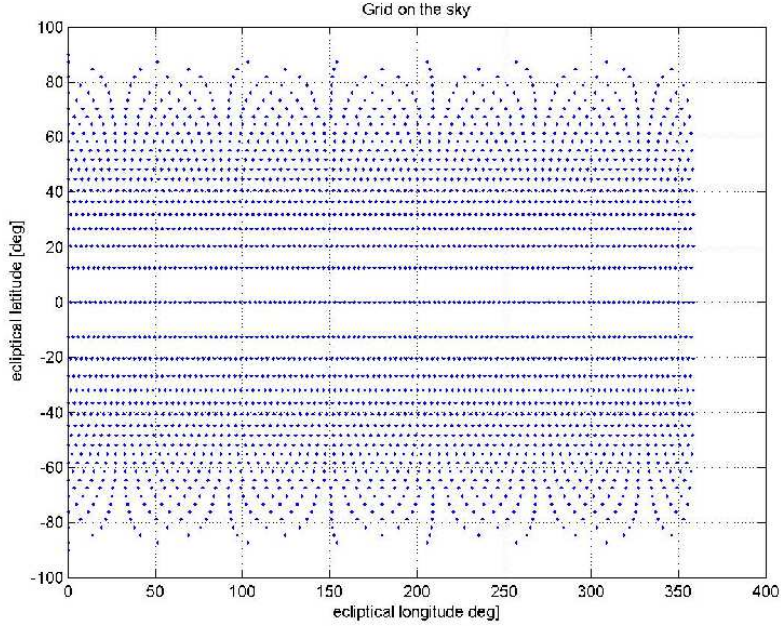}
\caption{\small \sf 
Optimal sky map, for $N_D=20$, $K_{sky}=1$.
x-axis: Ecliptical longitude, degrees, from 0 to 400;
y-axis: Ecliptical latitude, degrees, from -100 to 100;  
The number of points in the map is $N_{sky}$=2902.}  
\label{fig:gridsim1}
\end{center}
\end{figure}

Here we describe how we construct the grid on the sky.
Suppose two sources, at the same frequency $f_0$ and same latitude $\beta$.
Their angular delay $\gamma$ with respect to the detector rotation
produces a time delay $\Delta t=\gamma/\Omega_{orb}$. The two sources will then
have the same frequency variation at the detector, which is the classical 
equation due to the Doppler effect, 
$$\Delta f=f_0 \frac{v_{orb} \cos \beta }{c} $$
but with time delay
$\Delta t$. 
The observed frequency difference has thus a maximum value which is
given by
\begin{equation}
\Delta f= 2 f_0 \frac{v_{orb} \gamma \cos \beta }{c}
\end{equation}
Thus  the angular resolution is, in radians:
\begin{equation}
\gamma= 1/(N_D \, \cos \beta)
\label{gammalong}
\end{equation}
where
$N_D$ is the number of points in the
Doppler band for a signal of max frequency $f_0$:
\begin{equation}
N_D=2 \, f_0 \, v_{orb}/{\large( } c \,\Delta f {\large )}
\end{equation}
and
 $v_{orb}/c \approx 10^{-4}$. \\ 
We now repeat the same reasoning, supposing
the two sources, at the same frequency $f_0$ and same longitude $\lambda$.
The two sources will 
have the same frequency variation at the detector, now given by
$\frac{df}{d \beta}$,
but with an angular delay $\gamma'$. 
The observed frequency difference has a maximum value which is:
\begin{equation}
\Delta f= f_0 \frac{v_{orb} \gamma' \sin \beta }{c}
\end{equation}
We obtain for the angular resolution, in radians:
\begin{equation}
\gamma'= 1/(N_D \, \sin \beta)
\label{gammalat}
\end{equation}
Using  eqs. \ref{gammalong} and \ref{gammalat} we get:
\begin{equation}
\Delta \lambda= 1/(N_D \, \cos \beta)
\label{deltalong}
\end{equation}
\begin{equation}
\Delta \beta= 1/(N_D \, \sin \beta)
\label{deltalat}
\end{equation}
Using these equations we construct the grid on the sky, 
which we call the ``optimal'' grid.
The points of the grid are not uniformly distributed. 
With a simulation, we have estimated the
the number of points in the grid  $N_{sky}$, which is, in the high
frequency limit:
\begin{equation} 
N_{sky} \simeq K_{sky} \frac{2 \pi^2 N_D^2}{\pi} 
\end{equation}
$K_{sky}$ is an extra resolution factor, 
which can be
greater than 1, to enhance the efficiency, but even less than 1, to save
computing cost, obviously worsening the efficiency.
Fig.\ref{fig:gridsim1} shows the optimal sky grid, for $N_D=20$ 
(which corresponds to a source frequency $f_{0}= 100$ Hz).
As already said, the grid used in the sky Hough method, is not optimal, but
rectangular, to use fastest computing algorithms.
The number of points in this rectangular grid is:
\begin{equation} 
N_{sky}=K_{sky}\,2 \pi^2 N_D^2
\end{equation}
 which is, asymptotically, a factor $\pi$ higher then the number 
of points of the optimal grid. 
In fact this
grid has to be over resolved to maintain the same sensitivity of the
corresponding optimal grid. Further, we note that
this over resolution produces a higher number of candidates from certain
sky positions.
\begin{figure}
\begin{center}
\includegraphics[width=12cm]{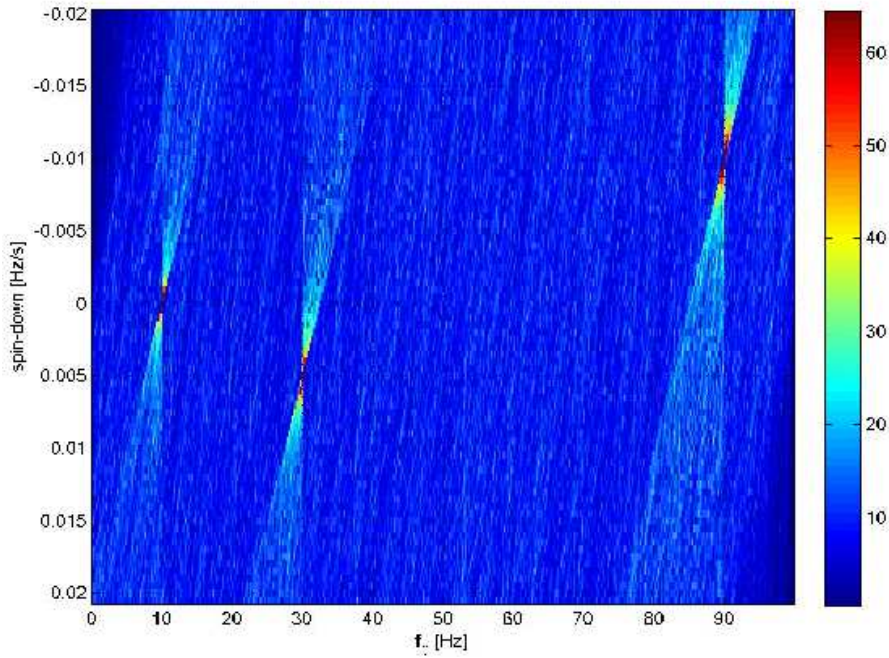}
\caption{\small \sf 
An example of a frequency Hough map, having injected
into white noise three signals, at different frequencies and spin-down. 
x-axis: Estimated source frequency, Hz, from 0 to 100;
y-axis: Spin down, Hz/s, from 0.02 to -0.02.}  
\label{fig:gridsim2}
\end{center}
\end{figure}
\subsection{Effect of artifacts}
The sensitivity of the sky Hough procedure is affected
by artifacts, i.e. an excess of candidates in
some places of the sky map, which are  
due to local spectral disturbances.
The effect can't
be eliminated because each map is constructed over the whole sky, and
hence the threshold for candidate selection has to be the same 
for the whole sky.
Using the frequency Hough procedure this effect disappears
because each map is constructed for only
one position in the sky. So, because of the adaptivity of the
threshold, if a sky region gives an excess of candidates, the threshold is
raised and then there  is a loss in sensitivity
only for that sky region. 
\section{Study the efficiency of the methods}
We are now ready to enter into details by studying the
efficiency of both the methods, by the use of simulations.
Figure \ref{fig:gridsim2} is an example of how a frequency Hough map 
looks like, having injected
into white noise three signals, at different frequencies and spin-down.
\subsection{Loss in amplitude vs frequency over resolution factor}
\begin{figure}
\begin{center}
\includegraphics[width=8cm]{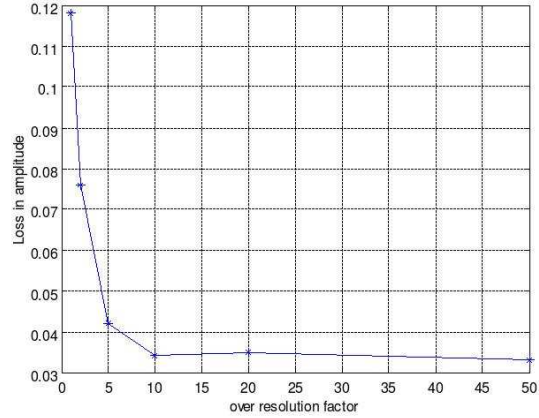}
\caption{\small \sf  
Loss in amplitude vs the frequency over-resolution factor.}  
\label{fig:freqloss}
\end{center}
\end{figure}

To study the efficiency of the methods, as a function of the frequency
over resolution factor, we have simulated a signal  
in the absence of noise. 
The reason for this is that we were interested in studying only the 
losses due to the discretization errors.
The parameters chosen for the simulation 
are similar to actual situations
(detector parameters, source expected parameters).
The parameters of the simulation are shown in Table \ref{tab:par}. 
Fig. \ref{fig:freqloss} shows the amplitude loss versus the frequency 
over resolution factor $r$. 
The loss was calculated as the average value of all the peaks found in the
500 spectra (it is important to remember that our procedure considers
peaks only the maxima above threshold).
The result is clear:
using $r=10$ the amplitude loss is 3.6 $\%$ (the efficiency $96.4 \%$),   
while with $r=1$, which is the only practically possible choice of 
the sky Hough, 
 the amplitude loss is 11.6 $\%$ (the efficiency $88.4 \%$).
From the figure, we notice that
there is no further gain of increasing the over resolution factor over 10.
Thus, we fixed to 10 the over resolution factor for the frequency Hough.
In next simulations, results with $r=10$ are thus for the frequency Hough,
results with $r=1$ are for the sky Hough.

\begin{figure}
\includegraphics[width=8cm]{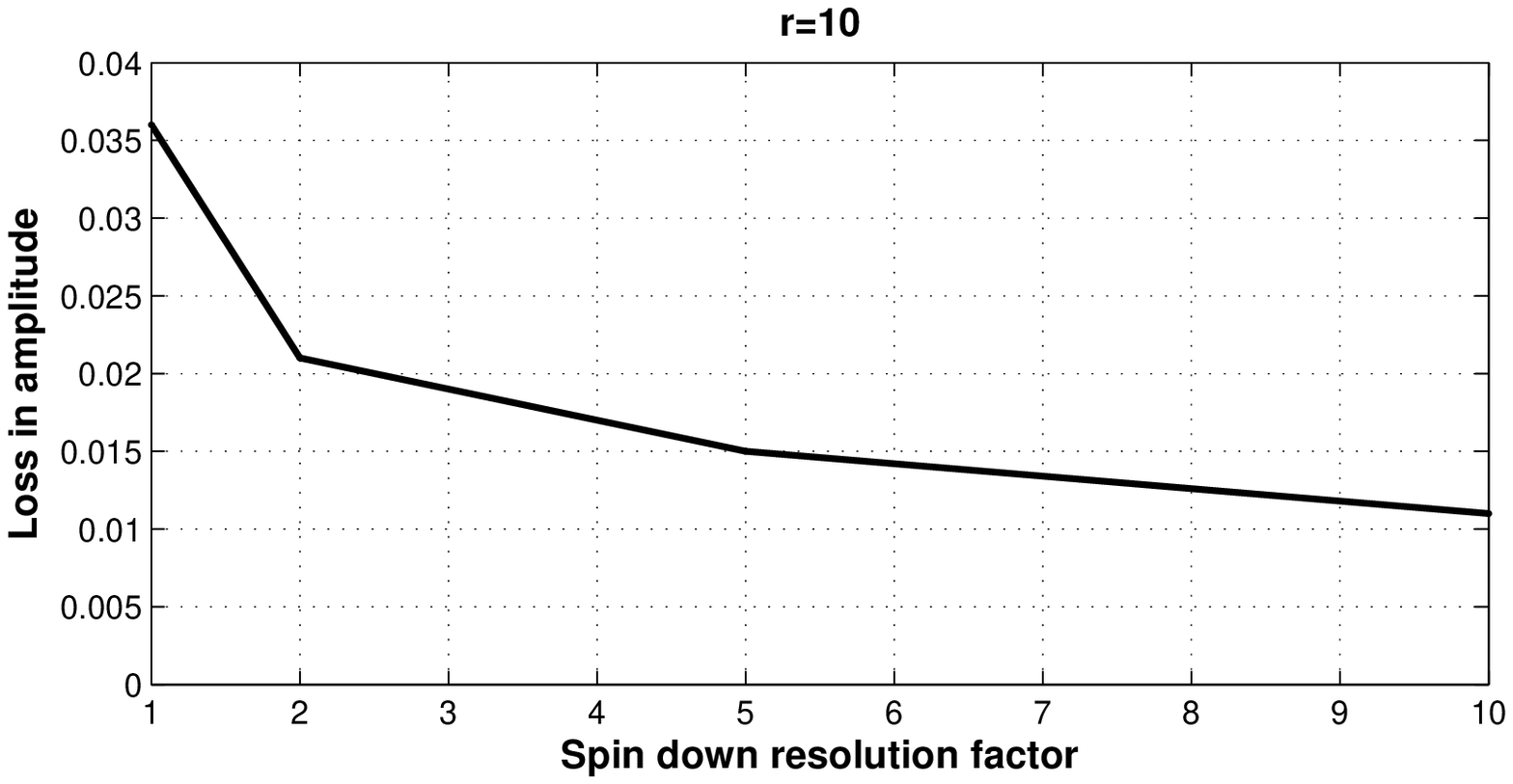}
\includegraphics[width=8cm]{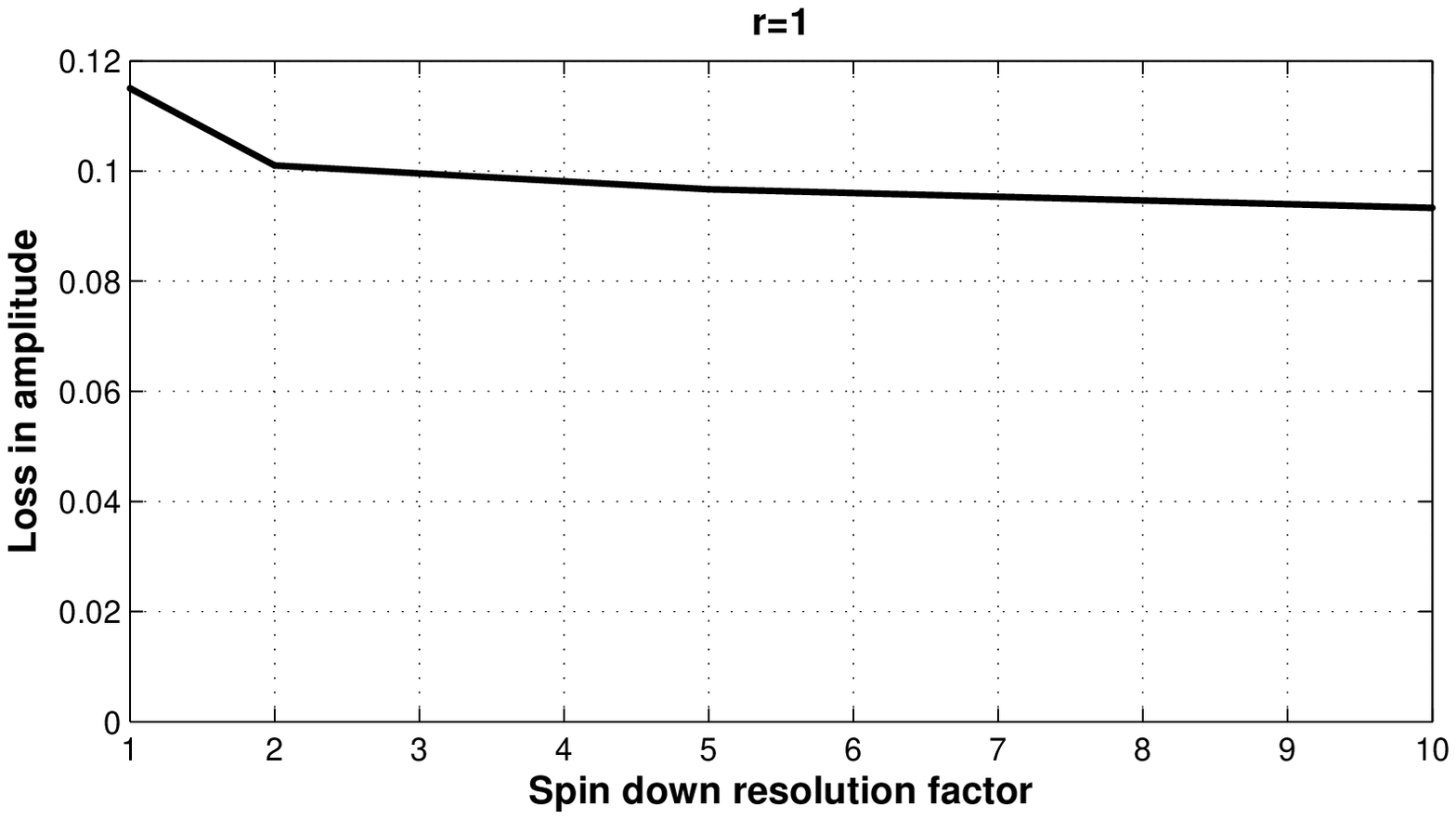}
\caption{\small \sf 
Loss in amplitude vs the spin down over-resolution factor,
for the two cases: $r=10$ (left) and $r=1$ (right).}
\label{fig:freqloss1}
\end{figure}
\subsection{Loss in amplitude vs the spin down resolution}
Once we have fixed the frequency over resolution factor 
we wanted to
quantify how the 
increasing of the spin down resolution from the nominal one would affect the
sensitivity. 
The results are  in Fig. \ref{fig:freqloss1}, which 
shows the loss in amplitude
vs the spin down over resolution factor, for both the cases $r=1$,
sky Hough, and $r=10$,frequency Hough. 
It can be noticed that, in the case of the frequency Hough,
even for the worst analyzed
situation, which corresponds to the nominal spin down step  
$$\Delta d=\frac{\Delta f}{T_{obs}}$$ 
the loss is quite small. Is is 
 3.6 $\%$ (the efficiency $96.4 \%$).
The situation is worst for the sky Hough, where the loss in amplitude at the
nominal spin down step is 11.6 $\%$ (the efficiency $88.4 \%$). The improvement
obtained by a better spin down resolution is not so important, as can be seen
from the figure.
It seems reasonable, given the
observation that increasing the spin down resolution has a computational
cost for both the methods, 
to use the nominal $\Delta d$ resolution (x-axis equal to 1 in the figure).
\subsection{Loss in amplitude vs sky grid resolution}
\begin{figure}
\includegraphics[width=8cm]{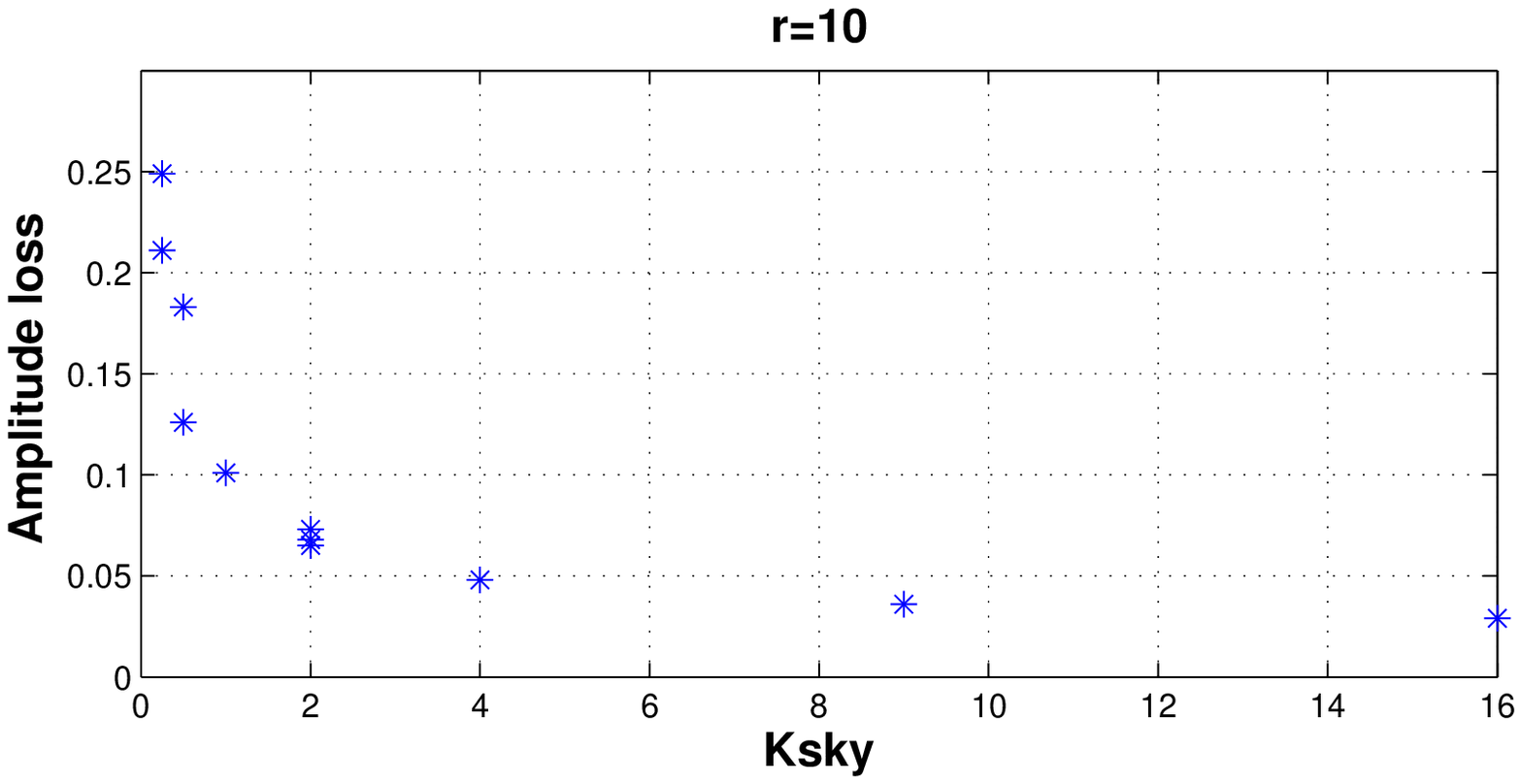}
\includegraphics[width=8cm]{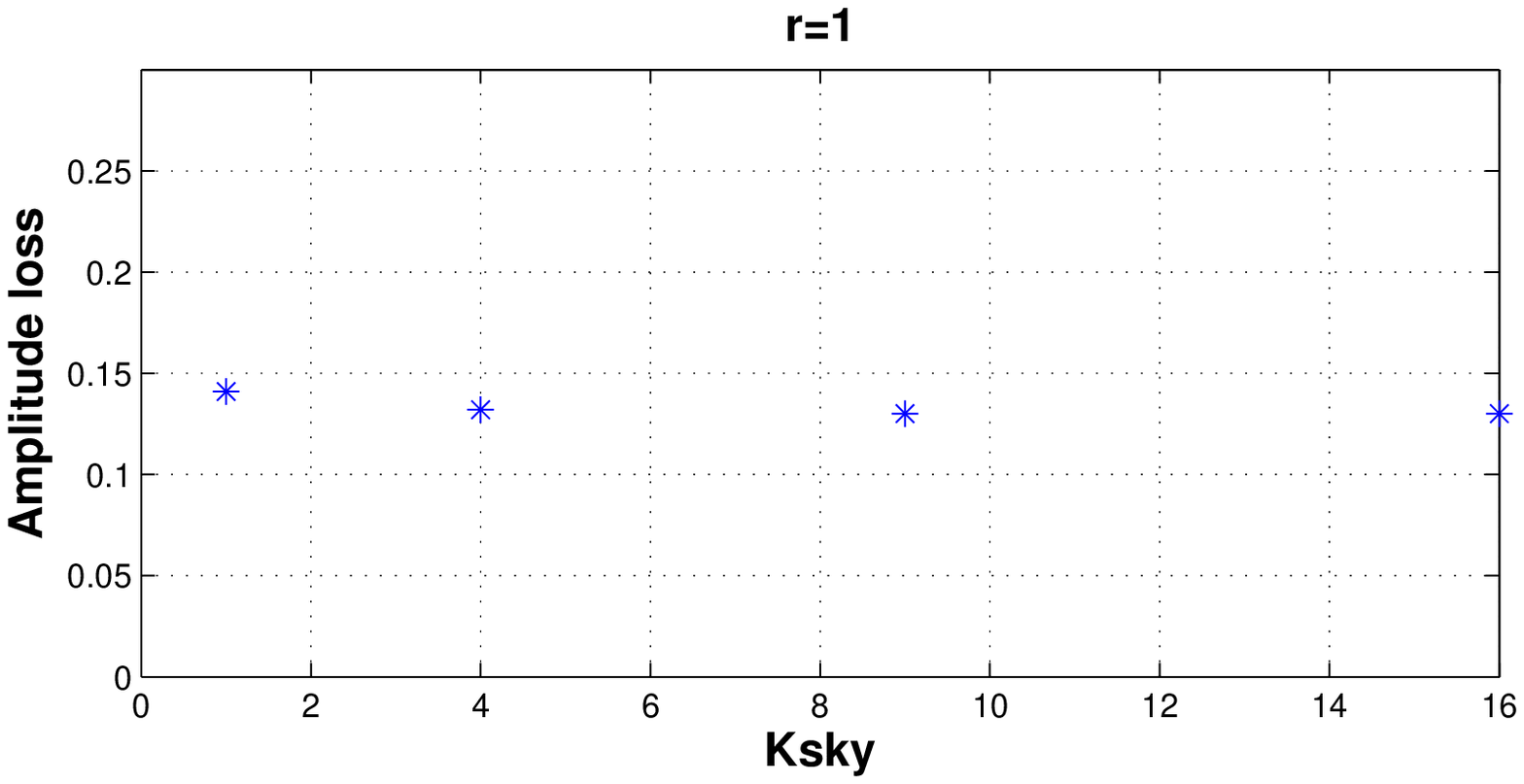}
\caption{\small \sf Loss in amplitude versus the sky map resolution factor $K_{sky}$.
The figures compare the loss when $r=10$ (left), frequency Hough,
and when $ r=1$ (right), sky Hough.}  
\label{fig:loss_spinres}
\end{figure}

To study the loss due to the sky grid resolution, we have simulated 
50 signals, randomly distributed over the sky. We have then looked for
results using the optimal grid, again registering the average value of
all the detected peaks.
In what follows, we suppose to use the optimal grid for both the procedures,
sky and frequency Hough.
Fig.\ref{fig:loss_spinres} shows the amplitude losses, as a function
of the over resolution sky map factor $K_{sky}$, in the two cases of
$r=10$ (left), frequency Hough, and
$r=1$ (right), sky Hough.
The amplitude loss, for $K_{sky}=1$, is $10 \%$ for the frequency Hough,
and   $14 \%$, for the sky Hough.
 Again, a better efficiency for
the new procedure. We notice that the use of an over resolution for the sky map, would
have an impact on the computing cost, with both the procedures.
\begin{table}
\begin{center}
\begin{tabular}{|l|c|c|c|}
\hline
 $\Delta f$ [Hz] & Source frequency [Hz]  & d [Hz/s] & $\Delta d$ [Hz/s]\\
\hline
\hline
&&& \\
$10^{-3}$ & 50  & $5 \cdot 10^{-8}$   &  $2 \cdot 10^{-9}$  \\
&&& \\
\hline
\end{tabular}
\caption{Parameters used in the simulation. 
The number of SFDBs is 500. 
The nominal spin down resolution is $\Delta d=\frac{\Delta f}{T_{obs}}$.}
\label{tab:par}
\end{center}
\end{table}
\mbox{}\\
\section{Summary of the results and conclusions}
Results of this study are summarized in Table \ref{tab:final}
\begin{table}
\begin{center}
\begin{tabular}{|l|c|c|c|}
\hline
& Sky & Frequency and spin-down  & Total \\
\hline
\hline
&&& \\
Sky Hough method & 0.860  & 0.88  &  0.757 (squared 0.573) \\
&&& \\
\hline
&&& \\
Frequency Hough method & 0.900  & 0.964  &  0.868 (squared 0.753) \\
&&& \\
\hline
\end{tabular}
\caption{Comparison of $h$ efficiencies. The table reports, for both the
methods, the partial (for unknown sky position, and for unknown
 frequency and spin down values), and the total efficiencies.}
\label{tab:final}
\end{center}
\end{table}
\mbox{}\\
We see that the ratio of the amplitude efficiencies is
\begin{equation} 
\frac{freq.~Hough}{sky~Hough}= 1.148
\end{equation}
which in power is 1.317.
From this, we can compute the gain in computing cost
for the same sensitivity.
Let us firstly recall that the $h$ sensitivity in the 
hierarchical search is proportional to ${T_{coh}}^{\frac{1}{4}}$, and
the computing cost to  ${T_{coh}}^3$.
Thus, the ``equivalent  FFT'' length factor  is $1.148^4$=1.734 and the gain in
computing cost is $1.734^3$=5.2 
(that is, the ratio of computing costs needed to have the same $h$ sensitivity).

Let's conclude with two further considerations, regarding the characteristics
of the frequency Hough procedure:
\begin{itemize}
\item
the {\bf adaptivity}, that is the weight of peaks to consider the noise level 
and the gain due to the antenna pattern toward a direction,
is, with this approach, immediate and very simple, as each Hough map is done
for a single sky position.
it has been shown, with the sky Hough, that the adaptivity of the procedure
is a very important task for the analysis;
\item
this new procedure is  appropriate also for all those situations 
in which the {\bf source position} is known and we should 
estimate only source frequency and spin down; 
\item
with a proper choice of parameters, it is also possible to detect and hence remove {\bf spurious signals}, with a constant or linearly varying frequency.
\end{itemize}
On the latter point, we are 
now working to study the efficiency of this method in terms 
of rejection of spurious lines in the peakmap. We know that this
is a very critical task for the analysis, since 
the presence of spurious lines highly 
affects the sensitivity of the search. 
We expect this new method to be
much more insensitive to the presence of spurious lines, since in 
the chosen Hough plane spurious lines and g.w. signals should have a very
different and well separable behavior.

\end{document}